# Construction and calibration of a goniometer to measure contact angles and calculate the surface free energy in solids with uncertainty analysis

Jonathan M. Schuster[1,2,*], Carlos E. Schvezov[1], Mario R. Rosenberger[1]

[1]Instituto de Materiales de Misiones (IMAM), Universidad Nacional de Misiones-Consejo Nacional de Investigaciones Científicas y Técnicas, Posadas, Misiones, Argentina.

[2]Instituto Sabato, Universidad Nacional de San Martín-Comisión Nacional de Energía Atómica, San Martín, Buenos Aires Argentina.

*Corresponding author: jschuster@fceqyn.unam.edu.ar

**Abstract**

Here, we present the construction and calibration of a low-cost goniometer to measure contact angles by the sessile drop method. Besides, we propose a simple and fast method to calculate the uncertainty in the determination of the surface free energy (SFE) and its polar and dispersive components through the Owens-Wendt model and tested it by using two testing liquids. The goniometer performance and the SFE uncertainty were determined on two polymers: polytetrafluorethylene (PTFE) and polyoxymethylene (POM), by using water and methylene iodide. The values of contact angle measured were used to calculate the SFE and its components with their errors. The SFE values obtained for PTFE were 17.57-17.91 mJ/m$^2$, with a relative error lower than 5.5 %, whereas those for POM were 42.80-43.23 mJ/m$^2$, with a relative error lower than 4.3%. Both the SFE values and the errors were in the range of those previously reported. Based on the mathematical analysis of the uncertainty propagation in the determination of SFE, we concluded that the uncertainty is minimized when the testing liquids are an apolar liquid and water.

**Keywords:** Sessile drop, Goniometer, Experimental errors, Surface free energy, Calibration

**1. Introduction.**

The contact angle of liquids on solid surfaces is an important parameter in many industrial products such as medical devices, adhesives, paints, coatings and cosmetics [1]. In addition, the contact angle is used to calculate the surface free energy (SFE) of solids ($\gamma_S$) by means of different theoretical equations [2]–[6]. $\gamma_S$ is defined as the work necessary to apply in order to produce a new material surface, and is given by the energy/surface area relationship. In the MKS system, the units are mJ/m². At the atomic or molecular level, $\gamma_S$ represents the degree of difference between the forces at a solid-vapor interface and the forces inside the bulk solid [7].

One of the experimental methods to measure the contact angle is the sessile drop method. A sessile drop is a liquid drop sitting in contact with a solid surface. The shape of the drop forms an angle (θ) between the solid surface and the tangent line between the liquid drop and the ambient atmosphere (gas or vapor) starting at the triple point (solid-drop-atmosphere) towards the liquid phase, as shown in Figure 1. θ is defined as the contact angle. The ideal contact angle is that formed with a perfectly smooth, inert and chemically homogenous solid surface, in the absence of external fields such as gravity or electromagnetic. The value of the contact angle is the result of the physicochemical interactions between the three phases, resulting in an equilibrium of the intermolecular forces present, including the cohesive and adhesive forces between the liquid and the solid [1]. The value of the contact angle could be between 0 and 180 °.

The Young equation establishes a relation between the contact angle and the interfacial tensions, shown in Figure 1, as follows:

$$\gamma_{LV} \cos \theta = \gamma_{SV} \, \gamma_{SL} \tag{1}$$

where $\gamma_{SV}$, $\gamma_{SL}$ and $\gamma_{LV}$ are the interfacial tensions between solid-vapor, solid-liquid and liquid-vapor, respectively [8], [9].

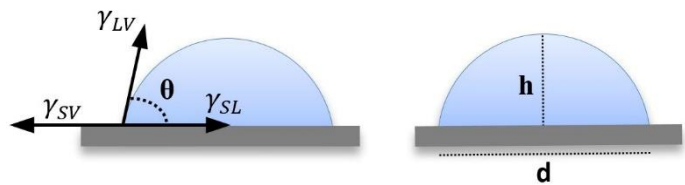

*Figure 1. Drop of a liquid on a solid surface.*

In the case of a real solid surface, the contact angle has a local value associated with the local properties of the solid, which may vary along the points of the line formed by the contact of the three phases (liquid, solid and vapor). In the case of a very homogenous and chemically inert solid surface with a low roughness, it may be assumed that there are no significant variations of the contact angle from point to point along the triple contact line, and that its value is a local average, considering that the possible existing differences are not detected due to the resolution of the optical observation and angle measurement [10].

Gravity exerts a force, which, in the case of very small drops of the order of microliters, is negligible compared with the adhesive and cohesive forces, and, therefore, it may be assumed that the drop shape is defined by these two main forces [11], [12]. The contact angle could be either static, if the contact line formed by the three phases remains steady, or dynamic, if the contact line moves back and forth during the measurement [13].

The calculation of SFE based on contact angle measurements has been used and reported as a standard and accepted method; however, the uncertainty associated with the calculated value of SFE has been partially analyzed [2]–[4], [6], [14], [15].

In the present report, we present and describe a goniometer to measure contact angles. The main motivation of the present work is to produce a goniometer at a much lower cost than commercial goniometers available in the market with similar characteristics, which could be built by research groups with a low research budget. The equipment was calibrated using procedures designed for this purpose. The results, as well as the possible occurrence of systematic and random errors, are presented and analyzed. The goniometer was used to measure the contact angles of drops of two liquids, water and methylene iodide, on two solids, polytetrafluorethylene (PTFE) and polyoxymethylene (POM). The values of the contact angles were used to calculate the SFE of PTFE and POM, using the geometric mean model. An easy method to calculate the error on the SFE value and its components due to the error on the contact angle value is presented and used. An analysis was made on how errors on the contact angle values of different liquid drops determine the value of the propagated errors on the SFE values.

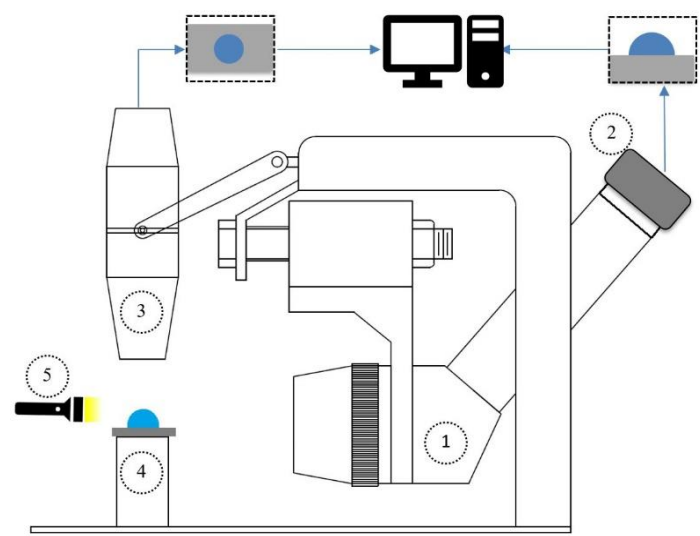

*Figure 2. Goniometer scheme. Ref.: 1) monocular microscope (Arcano DM1), 2) webcam (LifeCam HD-3000 Microsoft), 3) digital microscope (Generic Brand), 4) support for the sample, 5) diffuse light.*

## 2. Materials and methods

### 2.1. Goniometer

The goniometer is schematically represented in Figure 2. The main components are a monocular microscope (objective 2x, ocular 16x), a webcam and a digital microscope (up to 200x). The webcam is used to record the shape of the drop from a lateral position to measure the contact angle, whereas the microscope located above the drop is used to observe and analyze the drop symmetry from the top.

The basement and the column holding the monocular microscope and digital microscope is made of stainless steel AISI 304. The final total cost for the present goniometer was about 200 US dollars with a work load of 30 h.

### 2.2. Calibration standards

Stainless steel hemispheres, which were cut from full spheres, were used as calibration standards. The hemispheres were polished at the base up to 1 µm diamond paste in ethylene glycol to obtain a flat surface. The hemisphere height ($h$) was measured using a micrometer Schwyz (0-25 mm ± 0.001 mm). The diameter of the base of each hemisphere was calculated using equation 2, where R is the radius of the full sphere and $h$ is the hemisphere height:

$$d = 2\sqrt{2Rh - h^2} \quad (2)$$

The propagated errors in $d$ due to the errors in R and $h$ were calculated using equation 3 [16]:

$$E_d = \sqrt{\left(\partial d/\partial R|_{R,h}\right)^2 (E_R)^2 + \left(\partial d/\partial h|_{R,h}\right)^2 (E_h)^2} \quad (3)$$

where $E_i$ is the absolute error of variable i=d,R,h and $\partial d/\partial i|_{i,j}$ is the derivate of $d$ from equation 2 with respect to variable i, at the average measured value of variables i and j, i,j=R,h. The result for $d$ is reported as $d = \bar{d} \pm E_d$.

The value of θ was calculated using the values of $d$ and $h$ determined previously, using the following equation 4 [11], [17]:

$$\theta = 2\tan^{-1}(2h/d) \quad (4)$$

The error in θ is calculated in similar way:

$$E_\theta = \sqrt{\left(\partial\theta/\partial h|_{d,h}\right)^2 (E_h)^2 + \left(\partial\theta/\partial d|_{d,h}\right)^2 (E_d)^2} \qquad (5)$$

where $E_h$ and $E_d$ are the absolute errors in *d* and *h* obtained previously. The result is reported as $\theta = \bar{\theta} \pm E_\theta$.

### 2.3. Optical distortion

Considering that any optical instrument is subject of some degree of distortion, it is necessary to determine the type and degree of distortion to correct the measurements made from the optical images and to develop a reliable contact angle measurement procedure.

The three possible distortions in this case are barrel (negative radial distortion), pincushion (positive radial distortion) and mustache distortion (a combination of the barrel and pincushion distortions) [18]. The three possible distortions are illustrated in Figure 3.

The procedure to detect, quantify and correct any of the above distortions consists in putting three glass slides piled up one on top of the other and obtaining an image with the experimental optical system conformed by the webcam and the monocular microscope.

The pictures are shown in Figure 3 (d) and (e) from the lateral view of the piled-up slides rotated 90º one from the other, respectively. In each picture, three distances were measured: vL1, vL2 and vL3 in one case, and hL1, hL2 and hL3 in the other case, as illustrated in Figure 3 d) and e) respectively. The presence of pincushion distortion is detected if vL1≅vL3>vL2 and hL1≅hL3>hL2, whereas the presence of barrel distortion is detected if L1≅vL3<vL2 and hL1≅hL3<hL2. In the absence of distortion, vL1≅vL3≅vL2≅hL1≅hL3≅hL2. Any distortion detected in the system was corrected with the software ShiftN 4.0 [19].

### 2.4. Measurement of the standards.

Each steel hemisphere standard resembles a regular liquid drop placed on a slide and was photographed using a picture system. A total of 10 pictures for each of the three hemispheres were taken in different places of the slide. Each picture was processed and analyzed with the software ImageJ [20] to measure the contact angle. The following three different procedures were used to determine the contact angle.

#### 2.4.1. Procedure I.

Using the software ImageJ, the 10 digital pictures of the standard drop were used to measure *d* and *h* of the drop and to calculate the contact angle from equation 4.

#### 2.4.2. Procedure II

In this procedure, no assumption was made about the drop shape or symmetry. The section of the standard drop was represented from 15 to 20 points selected from the image of the drop. The shape of the drop was used to determine the contact angle for each side, i.e. the left and the right, of the image. The reported values are the arithmetic mean of both angle values. In this case, the plug-in, Drop Analysis–DropSnake from the software ImageJ [21] was used.

### 2.4.3. Procedure III

In this case, the drop was assumed to be axisymmetric, although it may not be the case. The procedure corrects for the possibility of drop deformation due to gravity, and the drop shape is adjusted using the Young-Laplace equation and the value of θ for each drop image is obtained using the plugin, Drop Analysis – LB-ADSA from ImageJ software [22].

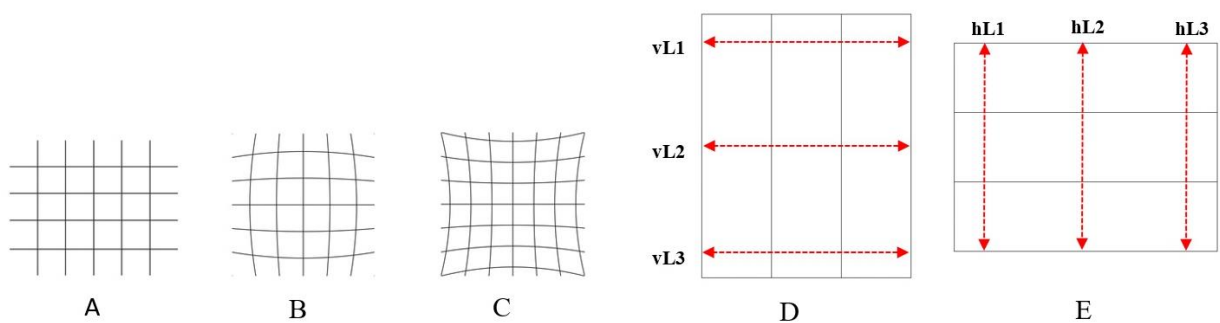

*Figure 3. Optical distortion: (A) without distortion; (B) barrel (negative radial distortion); (C) pincushion (positive radial distortion). Piled-up slides: (D) vertical and (E) horizontal.*

### 2.5. Contact angle measurement of water and methylene iodide drops on PTFE and POM.

The solid surface of PTFE was polished using SiC papers of grit sizes #320 to #1500 and then with 1 µm diamond paste lubricated with ethylene glycol for 20 minutes at 60 rpm. The solid surface of POM was polished using SiC papers of grit sizes #500 to #1500 and then placed between two glass slides and kept for 10 minutes in a furnace at 180 ºC. This allowed obtaining a good smooth surface.

The surface roughness was determined using a profilometer TR200 with RC phase contrast filter and a cut-off of 0.8 mm on a 4-mm long line. The average roughness values from five measurements for each solid were $Ra = 123\pm40$ nm and $Rq=172\pm59$ nm for PTFE and $Ra = 269\pm39$ nm and $Rq=367\pm92$ nm for POM.

Before depositing a drop to measure the contact angle, the PTFE and POM surfaces were cleaned with detergent and water, rinsed with ethanol, dried in hot air, then cleaned with acetone and dried again with hot air.

Immediately after, the sample was placed in the goniometer and a drop of liquid carefully deposited on its surface by using a micropipette. The liquids used in the experiments were deionized water (w) and methylene iodide (m). The drop volume was 1 µL with negligible effect of gravity. A total of eight drops were measured for each solid surface and liquid. The ambient temperature and humidity were 20±3 °C and 50±5 %, respectively. The contact angle was measured using the three procedures described above and the arithmetic mean, standard deviation (SD) and standard error of the mean (SEM) were calculated from the eight values of contact angle determined with each procedure. The value of contact angle is reported as the arithmetic mean±error, i.e. $\bar{x} \pm E_{CA}$, where $E_{CA}$ is three times the SEM.

Measurements with values outside the arithmetic mean ± three times the SD were discarded, and the new arithmetic mean and deviation were calculated and reported. This procedure was applied only once [16], [23]. In case there was more than one value outside these limits, more measurements were performed since the scatter may be the result of a high chemical or physical heterogeneous material, due to intrinsic properties of the solid, contamination or abnormalities in the drop deposition process.

### 2.6. Determination of SFE.

The values of contact angles of water ($\theta_w$) and methylene iodide ($\theta_m$) of drops deposited on POM and PTFE were then used to calculate the SFE of the solids ($\gamma_S$) by using the geometric mean (GM) approach [4]. In equation 1, the diffusion pressure ($\pi_e = \gamma_S - \gamma_{SV}$), which is the reduction in surface tension due to vapor adsorption, is neglected; in such case, $\gamma_{sv} \cong \gamma_s$. According to the GM model, the total SFE $\gamma_S$ of a substance i comes from two components: the dispersive and polar components:

$$\gamma_i = \gamma_i^d + \gamma_i^p \tag{6}$$

According to the GM method, the interfacial solid/liquid energy can be evaluated using the following equation [4]:

$$\gamma_{SL} = \gamma_S + \gamma_L - 2\left(\sqrt{\gamma_S^d \gamma_L^d} + \sqrt{\gamma_S^p \gamma_L^p}\right) \tag{7}$$

which, combined with equation 1, results in:

$$\gamma_L(1 + \cos\theta) = 2\sqrt{\gamma_S^d \gamma_L^d} + 2\sqrt{\gamma_S^p \gamma_L^p} \tag{8}$$

To calculate $\gamma_S$ by equation (6), the values of $\gamma_i^d$ and $\gamma_i^p$ are necessary. To calculate $\gamma_S^d$ and $\gamma_S^p$, equation 8 is linearized as:

$$\frac{0.5\,\gamma_L\,(1+\cos\theta)}{\sqrt{\gamma_L^d}} = \sqrt{\gamma_S^p}\left(\frac{\gamma_L^p}{\gamma_L^d}\right)^{1/2} + \sqrt{\gamma_S^d} \tag{9}$$

where $\sqrt{\gamma_S^p}$ and $\sqrt{\gamma_S^d}$ are the slope and the ordinate to the origin respectively in a linear plot, where the variables $0.5\,\gamma_L(1+\cos\theta)/\sqrt{\gamma_L^d}$ and $\sqrt{\gamma_L^p/\gamma_L^d}$ are the ordinate and the abscissa, respectively. The minimum number of points to obtain a solution is two, which is obtained using only two different liquids. The calculated values of surface tensions and their components for water and methylene iodide are listed in Table 1.

### 2.7. Estimation of the maximum error of $\gamma_S^p$, $\gamma_S^d$ and $\gamma_S$

The values of the contact angles of water ($\theta_w \pm E_{CA}$) and methylene iodide ($\theta_m \pm E_{CA}$) were then used to draw two linear functions following equation 9, with the values of ($\theta_w + E_{CA}$) and ($\theta_m - E_{CA}$) in the first case and with those of ($\theta_w - E_{CA}$) and ($\theta_m + E_{CA}$) in the second case. In such case, from these linear functions, two values for $\gamma_S^p$ ($\gamma_S^p{}_{max}$ and $\gamma_S^p{}_{min}$) and $\gamma_S^d$ ($\gamma_S^d{}_{max}$ and $\gamma_S^d{}_{min}$) were obtained and averaged to give:

$$\overline{\gamma_S}^p = (\gamma_S^p{}_{max} + \gamma_S^p{}_{min})/2 \tag{10}$$

$$\overline{\gamma_S}^d = (\gamma_S^d{}_{max} + \gamma_S^d{}_{min})/2 \tag{11}$$

The absolute errors associated with each polar ($E_{\overline{\gamma_S}^p}$) and dispersive ($E_{\overline{\gamma_S}^d}$) component are:

$$E_{\overline{\gamma_S}^p} = (\gamma_S^p{}_{max} - \gamma_S^p{}_{min})/2 \tag{12}$$

$$E_{\overline{\gamma_S}^d} = (\gamma_S^d{}_{max} - \gamma_S^d{}_{min})/2 \tag{13}$$

Finally, using equation 6, the total SFE value and its respective absolute error is:

$$\overline{\gamma_S} \pm E_{\overline{\gamma_S}} = \left[\overline{\gamma_S}^p + \overline{\gamma_S}^d\right] \pm \left[E_{\overline{\gamma_S}^p} + E_{\overline{\gamma_S}^d}\right] \tag{14}$$

Based on the error analysis developed in Appendixes A and B, it results that the use of water (w) and methylene iodide (m) to measure contact angles in solid surfaces will produce the lowest error in the calculation of the value of SFE since the liquids satisfy the required conditions resulting from the analysis, which are that the liquids have different values of the ratio between polar and dispersive components and that one of the liquids is apolar.

*Table 1. Data of surface tension and components of the test liquids used in this work* [2].

| Liquid | Purity | $\gamma_L$ [mJ/m²] | $\gamma_L^d$ [mJ/m²] | $\gamma_L^p$ [mJ/m²] | Polarity [$\gamma_L^p/\gamma_L^d$] |
|---|---|---|---|---|---|
| Water (w) | Deionized | 72.8 | 21.8 | 51.0 | 2.3 |
| Methylene Iodide (m) | ReagentPlus®, 99% | 50.8 | 50.8 | 0 | 0 |

## 2.8. Data processing

Each set of raw results was processed in the following way. First, the arithmetic mean values were calculated and then the SD and the standard error ($SE=SD/n^{1/2}$) determined. The data were also tested for normality using the modified Shapiro-Wilks test and the Q-Q plot. Finally, a 99% confidence interval was generated around the arithmetic mean by using the t-Student test. The results are presented in the form of a box-plot diagram and also as a point plot by using the software InfoStat 2016 in both cases. Accuracy controls of the goniometer and the procedure were carried out by checking each time that the contact angle of the standards was within the 99% confidence interval of the measurement.

## 3. Results and discussion

### 3.1. Standard measurements

Four standards consisting of hemispheres produced from steel balls were used as standard drops. The resulting contact angles measured with the procedures explained before are shown in Table 2. It is observed that the contact angles ranged from 37.19° to 119.66°, that the absolute error was constant and low (0.06°), and that the largest relative error was below 0.2% in all cases.

*Table 2. Contact angle of calibration standards.*

| Standards | R [mm] | h [mm] | $E_h$ [mm] | r [mm] | $E_r$ [mm] | θ [rad] | $E_\theta$ [rad] | θ [°] | $E_\theta$ [°] |
|---|---|---|---|---|---|---|---|---|---|
| A | 3.171 | 0.645 | 0.001 | 1.917 | 0.001 | 0.649 | 0.001 | 37.19 | 0.06 |
| B | 1.979 | 0.871 | 0.001 | 1.640 | 0.001 | 0.977 | 0.001 | 55.96 | 0.06 |
| C | 1.979 | 1.871 | 0.001 | 1.976 | 0.002 | 1.516 | 0.001 | 86.87 | 0.06 |
| D | 1.587 | 2.372 | 0.001 | 1.379 | 0.002 | 2.089 | 0.001 | 119.66 | 0.06 |

### 3.2. Optical distortion

The measurement of the lateral distances vL1, vL2, vL3, hL1, hL2 and hL3 for the pile up of three glass slides showed that the optical system produced a considerable pincushion distortion since vL1≅vL3>vL2 and hL1≅hL3>hL2. This distortion was corrected manually using the software ShiftN 4.0. The correction factors used were from -0.2 to -0.7. After applying each correction factor, the linear measurement of vL1, vL2, vL3 and hL1, hL2, hL3 was repeated and, as a result, it was found that using a factor of -0.65 the lengths measured were between 555 and 557 pixels (Figure 4), which resulted very satisfactory. Therefore, a correcting factor value of -0.65 was adopted to correct the pincushion distortion in the sessile drop measurements made with the present goniometer.

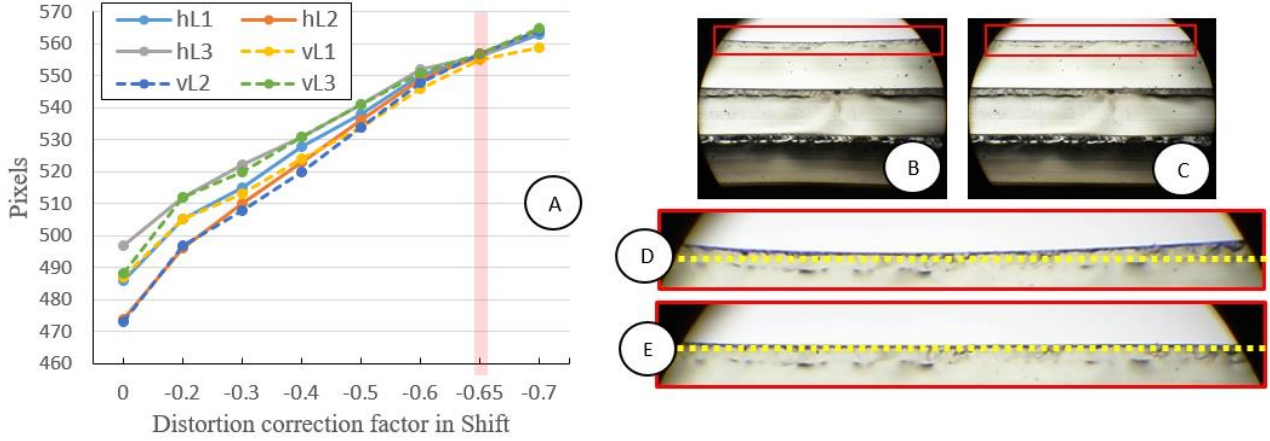

*Figure 4. (A) Dimensions vL1, vL2, vL3, hL1, hL2 and hL3 as a function of the correction factor applied to the image using the software ShiftN. Pictures of three slides piled up placed horizontally without correction (B); and with a correction using a correction factor of -0.65 (C). (D) Magnification of picture B; and (E) magnification of picture C.*

### 3.3. Contact angle measurements

#### 3.3.1. Procedure I

The distances $d$ and $h$ were measured in each image and these values were used to calculate the contact angles by using equation 4. The reported values of contact angle are the arithmetic mean calculated from 10 values obtained from 10 different images of the same drop. The results obtained for the four standard steel drops shown in Figure 5 are listed in Table 3.

The SD of the results decreased as the contact angle increased: 0.47 for standard A, with θ=37.4, and 0.22 for standard D, with θ=119.54; that is, an increase of 219 % in the value of contact angle produced a decrease of 113% in the deviation. Procedure I was therefore more precise for larger contact angles. This is attributed to the value of the condition number of equation 4 normally defined as [24]:

$$C(w) = \left| w \left[ \frac{d\theta}{dw} / \theta(w) \right] \right| \tag{15}$$

where $\theta$ is given by equation 4 and $w = 2h/d$. If $C(w) < 1$, $\theta(w)$ the function is well conditioned, $C(w) = 1$ the error in $\theta$ is the same as in $w$ and $C(w) > 1$ $\theta(w)$ is bad conditioned and the error in $w$ is magnified in $\theta(w)$.

From equations 4 and 15:

$$C(w) = |2w/[(1 + 4w^2)\tan^{-1}(2w)]| \tag{16}$$

As can be seen in equation 16, $C(w)$ is a decreasing function in the interval $(0, \infty)$ and is always lower than 1, that is, is a well-conditioned function, and the condition number for contact angles 37º, 56º, 87º and 120º are 0.93, 0.84, 0.66 and 0.41 respectively, showing an important decrease with an increase in contact angle. The absolute error in $\theta$ is always lower than 0.2º and the relative error decreases from 0.6% for standard A to 0.04% for standard B.

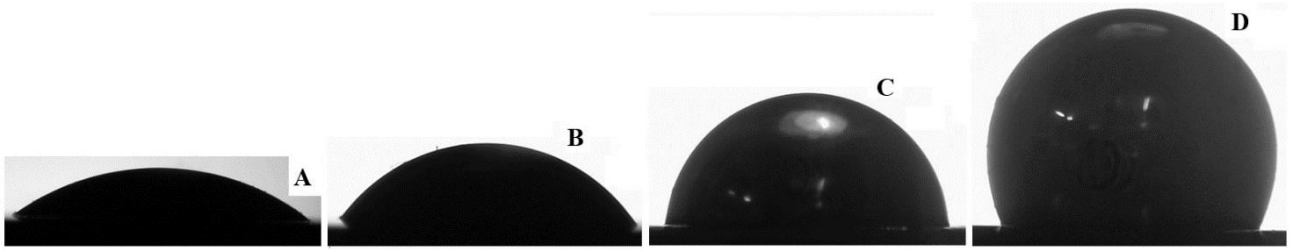

*Figure 5. Images of calibration standards A, B, C and D.*

### 3.3.2. Procedure II

The results using this procedure for the four standard steel drops were obtained as the arithmetic mean of 10 values of contact angle determined from each of the 10 images, as before.

The results are listed in Table 3. In contrast to the results obtained with Procedure I, there was no significant change in the SD of the arithmetic mean; i.e. 0.99 for standard A and 1.15 for standard D. In this procedure, as before, the absolute error did not exceed 0.2º in any case and the relative errors were low, from 0.5% for standard A to 0.1% for standards C and D.

### 3.3.3. Procedure III

As in procedures I and II, the reported value is the arithmetic mean of 10 contact angles determined from 10 different images of each standard.

The results shown in Table 3 indicate that the SD in the measurement was independent of the magnitude of the contact angle of the four standard steel drops. The absolute error was in all the cases lower than 0.1º and therefore the relative error decreased from 0.3% for standard A to 0.03% for standard C.

### 3.3.4. Analysis and comparison among Procedures

The results for the three procedures used to determine the contact angle in standard steel drops showed a normal distribution of data, within a confidence interval of 99% of the arithmetic mean (Table 4). Also, the box-plots of the results shown in Figures 6 and 7 are, in all cases, within the confidence intervals for the three procedures. For all the patrons and procedures employed the error is less or equal to 0.2°, which is in the lowest range of errors of commercial goniometers with similar characteristics reported to be between 0.2° to 1° [25]–[29].

*Table 3. Results of contact angle measurements.*

| Procedure | Calibration standards | Measure (n=10) [°] | | | $E_{CA}$ [°] | $\varepsilon_{CA}$ [%] |
|---|---|---|---|---|---|---|
| | | $\bar{x}$ | SD | SEM | | |
| I | A | 37.4 | 0.47 | 0.1 | 0.2 | 0.6 |
| | B | 55.98 | 0.20 | 0.06 | 0.02 | 0.04 |
| | C | 86.7 | 0.35 | 0.1 | 0.2 | 0.2 |
| | D | 119.54 | 0.22 | 0.07 | 0.12 | 0.10 |
| II | A | 37.4 | 0.99 | 0.3 | 0.2 | 0.5 |
| | B | 55.8 | 0.84 | 0.3 | 0.2 | 0.3 |
| | C | 87.0 | 0.91 | 0.3 | 0.1 | 0.1 |
| | D | 119.8 | 1.15 | 0.4 | 0.1 | 0.1 |
| III | A | 37.3 | 0.38 | 0.1 | 0.1 | 0.3 |
| | B | 56.1 | 0.34 | 0.1 | 0.1 | 0.2 |
| | C | 86.89 | 0.23 | 0.07 | 0.02 | 0.03 |
| | D | 119.7 | 0.35 | 0.1 | 0.1 | 0.1 |

$\bar{x}$: arithmetic mean, SD: standard deviation, SEM: standard error of the mean, $E_{CA}$: absolute error, $\varepsilon_{CA}$: relative error.

*Table 4. Interval of confidence and normality of the measurements by both procedures.*

| Procedure | Calibration standards | Normal distribution | | Confidence interval of the mean (99%) [°] |
|---|---|---|---|---|
| | | S-W (p-value) | Q-Q (r) | |
| I | A | 0.753 | 0.984 | [36.89 – 37.85] |
| | B | 0.023 | 0.934 | [55.78 – 56.18] |
| | C | 0.021 | 0.921 | [86.35 – 87.09] |
| | D | 0.233 | 0.954 | [119.25 – 119.79] |
| II | A | 0.149 | 0.949 | [36.36 – 38.40] |
| | B | 0.533 | 0.974 | [54.97 – 56.69] |
| | C | 0.989 | 0.992 | [86.11 – 87.95] |
| | D | 0.535 | 0.979 | [118.63 – 121.01] |
| III | A | 0.491 | 0.946 | [36.92 – 37.68] |
| | B | 0.603 | 0.957 | [55.80 – 56.13] |
| | C | 0.083 | 0.938 | [86.76 – 87.11] |
| | D | 0.505 | 0.969 | [119.38 – 120.10] |

S-W: Shapiro–Wilk test (p>0,01 normality of data is accepted), Q-Q: Q-Q plot (value of r close to 1 indicates Normal distribution of the data).

### 3.4. Contact angle of water and methylene iodide drops on PTFE and POM

The contact angles for water and methylene iodide drops sitting on smooth surfaces of PTFE and POM were determined with the three procedures described above. The results, which are shown in Table 5, are the average value of eight measurements for each procedure. The error reported is three times the SEM for which only one significant digit was used. Any two results within this interval were considered equal.

*Table 5. Results of contact angle measurements on PTFE and POM.*

| Procedure | POM | | | | | | | | | |
|---|---|---|---|---|---|---|---|---|---|---|
| | Water CA [º] | | | | | Methylene Iodide CA [º] | | | | |
| | $\bar{\theta}$ | SD | SEM | $E_{CA}$ | $\varepsilon_{CA}$% | $\bar{\theta}$ | SD | SEM | $E_{CA}$ | $\varepsilon_{CA}$% |
| I | 77.3 | 1.7 | 0.6 | 1.8 | 2.3 | 42.1 | 1.3 | 0.5 | 1.5 | 3.6 |
| II | 77.8 | 1.7 | 0.6 | 1.8 | 2.3 | 43.1 | 1.6 | 0.6 | 1.8 | 4.2 |
| III | 77.5 | 1.7 | 0.6 | 1.8 | 2.3 | 42.8 | 1.6 | 0.6 | 1.8 | 4.2 |
| | PTFE | | | | | | | | | |
| | Water CA [º] | | | | | Methylene Iodide CA [º] | | | | |
| | $\bar{\theta}$ | SD | SEM | $E_{CA}$ | $\varepsilon_{CA}$% | $\bar{\theta}$ | SD | SEM | $E_{CA}$ | $\varepsilon_{CA}$% |
| I | 109.9 | 0.9 | 0.3 | 0.9 | 0.8 | 80.2 | 1.2 | 0.4 | 1.2 | 1.5 |
| II | 110.1 | 1.3 | 0.5 | 1.5 | 1.4 | 80.0 | 1.4 | 0.5 | 1.5 | 1.9 |
| III | 110.4 | 0.8 | 0.3 | 0.9 | 0.8 | 80.6 | 0.6 | 0.2 | 0.6 | 0.7 |

$\bar{\theta}$: arithmetic mean, SD: standard deviation, SEM: standard error of the mean, $E_{CA}$: absolute error=3*SEM, $\varepsilon_{CA}$: relative error.

The values of contact angle of water and methylene iodide drops on POM were between 77.3º and 77.8º for water and between 42.1º and 43.1º for methylene iodide. In all cases, the SD was lower than 1.7º. There were no significant differences in the precision of the results of the contact angle among the procedures. The reported values of contact angle for water on POM using the sessile drop method are 73.5 [34], 75.9° [32], 77±2º [30], 78±2º [31] and 84.9 [33]. On the other hand, the values of contact angle for methylene iodide using the sessile drop ranged between 37.5° [35] and 55.5° [34]. These are showing that there is not a unique value of contact angle but a range of values, the contacts angles obtained here are in between the reported ones.

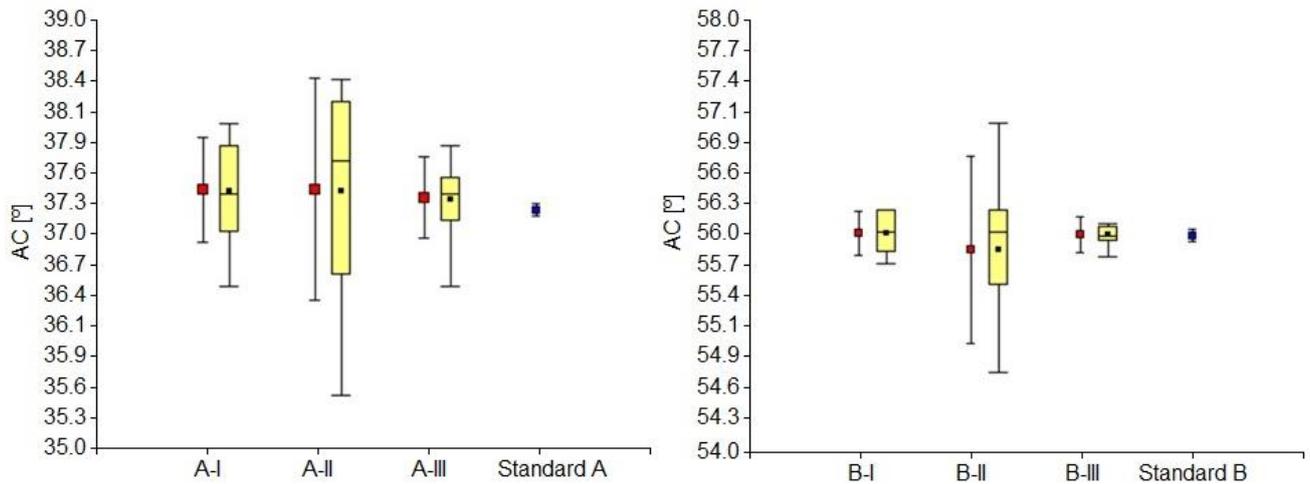

*Figure 6. Box plot and confidence interval of 99% for the contact angles determined by the three procedures (I, II and III) for drops A and B.*

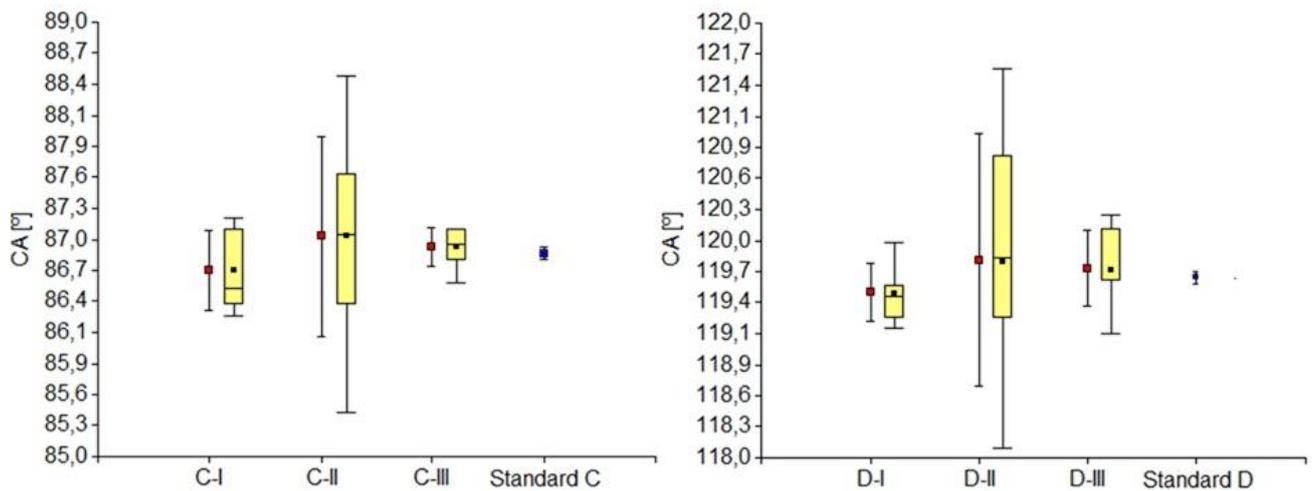

*Figure 7. Box plot and confidence interval of 99% for the contact angles determined by the three procedures (I, II and III) for drops C and D.*

In the case of water drops on PTFE, the measured contact angles were between 109.9º and 110.4º, with no significant differences in the SD among the procedures, whereas in the case of methylene iodide drops, the contact angles were between 80.0º and 80.6º, with a SD lower than 1.3º. The reported values of contact angle for water on PTFE using the sessile drop method ranged between 100° [36] and 121° [37]. On the other hand, the values of contact angle for methylene iodide using the sessile drop are 74.7° [38], 88° [39] and 91° [34]. The contacts angles obtained here are in between the reported ones.

### 3.5. Surface free energy of a PTFE and POM.

The values of SFE calculated using the geometric mean method and using the values of contact angles for water and methylene iodide determined by means of the three procedures described above (see Figure 8) are listed in Table 6.

In the case of POM, the values of SFE as well as its components and the errors were all similar. The polar component was ~4.7 mJ/m$^2$ ±0.9 mJ/m$^2$ (18% of relative error), whereas the dispersive component was ~38 mJ/m$^2$ ±0.9 mJ/m$^2$ (~2% of relative error). The total SFE was then ~43 mJ/m$^2$ ±1.8 mJ/m$^2$ (4% of relative error). Previously reported values of SFE of POM using the same liquids and the geometric mean method are of 1.62 mJ/m$^2$ for the polar component, 40.8 mJ/m$^2$ for the dispersive component, and 42±2.5 mJ/m$^2$ for the total SFE [40]. Another previous report in which the author did not specify the method or liquids used in the determination [41] showed a value of 14.1 mJ/m$^2$ for the polar component, 30.5 mJ/m$^2$ for the dispersive component, and 44.6 mJ/m$^2$ for the total SFE.

*Table 6. Surface free energy of a POM and PTFE.*

| POM | $\theta_w$ [°] | $\theta_{mi}$ [°] | $\gamma_S^p$ [mJ/m$^2$] | $\gamma_S^d$ [mJ/m$^2$] | $\overline{\gamma_S}^p$ [mJ/m$^2$] | $E_{\overline{\gamma_S}^p}$ [mJ/m$^2$] | $\varepsilon_{\overline{\gamma_S}^p}$ [%] | $\overline{\gamma_S}^d$ [mJ/m$^2$] | $E_{\overline{\gamma_S}^d}$ [mJ/m$^2$] | $\varepsilon_{\overline{\gamma_S}^d}$ [%] | $\overline{\gamma_S}$ [mJ/m$^2$] | $E_{\overline{\gamma_S}}$ [mJ/m$^2$] | $\varepsilon_{\overline{\gamma_S}}$ [%] |
|---|---|---|---|---|---|---|---|---|---|---|---|---|---|
| I | 79.1 | 40.6 | 3.85 | 39.31 | 4.70 | 0.85 | 18.1 | 38.53 | 0.78 | 2.0 | 43.23 | 1.63 | 3.8 |
|   | 75.5 | 43.6 | 5.55 | 37.75 |      |      |      |       |      |     |       |      |     |
| II | 79.2 | 41.3 | 3.89 | 38.95 | 4.79 | 0.90 | 18.8 | 38.01 | 0.94 | 2.5 | 42.80 | 1.84 | 4.3 |
|    | 75.6 | 44.9 | 5.68 | 37.06 |      |      |      |       |      |     |       |      |     |
| III | 79.3 | 41.3 | 3.85 | 38.95 | 4.71 | 0.85 | 18.0 | 38.17 | 0.78 | 2.0 | 42.88 | 1.64 | 3.8 |
|     | 75.7 | 44.3 | 5.56 | 37.38 |      |      |      |       |      |     |       |      |     |
| **PTFE** | | | | | | | | | | | | | |
| I | 110.8 | 79.0 | 0.26 | 18.00 | 0.42 | 0.16 | 38.1 | 17.40 | 0.61 | 3.5 | 17.82 | 0.77 | 4.3 |
|   | 109.0 | 81.4 | 0.58 | 16.78 |      |      |      |       |      |     |       |      |     |
| II | 111.6 | 78.5 | 0.18 | 18.27 | 0.41 | 0.23 | 56.1 | 17.50 | 0.77 | 4.4 | 17.91 | 0.99 | 5.5 |
|    | 108.6 | 81.5 | 0.64 | 16.73 |      |      |      |       |      |     |       |      |     |
| III | 111.3 | 80.0 | 0.26 | 17.49 | 0.38 | 0.12 | 31.6 | 17.19 | 0.31 | 1.8 | 17.57 | 0.43 | 2.4 |
|     | 109.5 | 81.2 | 0.50 | 16.88 |      |      |      |       |      |     |       |      |     |

Similarly, in the case of PTFE, the results were similar, independently of the procedure used: polar and dispersive components of ~0.4 mJ/m$^2$ ±0.15 mJ/m$^2$ (~40% of relative error) and ~17 mJ/m$^2$ ±0.8 mJ/m$^2$ (~4% of relative error), respectively and total SFE of ~17.9 mJ/m$^2$ ±0.9 mJ/m$^2$ (~4% of relative error). Previously reported values [4] using the same geometric mean method and the same liquids are of 0.5 mJ/m$^2$ for the polar component, 18.6 mJ/m$^2$ for the dispersive component, and 19.1 mJ/m$^2$ for the total SFE, which are within the experimental error of the values reported here. It is noted that the relative error in the value of the polar component is large (40%), a fact discussed below.

The values of absolute and relative errors of the SFE and its components obtained with the method proposed in this work are similar to those previously reported [14], [15].

## 4. Conclusions

A low-cost goniometer was built and calibrated to measure contact angles on solid surfaces by the sessile drop method. The goniometer is easy to operate, and the measured angles have high precision with a largest error of 0.2º and a maximum SD of 1.15º, comparable to the precision of commercial goniometers but at 10% of their cost. The goniometer was used to measure the contact angle of water and methylene iodide deposited surfaces of PTFE and POM to test the equipment and procedure. The images used to measure the contact angles were previously analyzed and the distortion of the system was corrected to minimize the error in the results.

The geometric mean method was applied to determine the SFE of solids from contact angle values. Previously, the full error of the procedure was analyzed, which indicated that, in order to reduce the error in the calculations of the SFE, the measurements of contact angles must be as precise as possible, the liquids used must have values of the ratios between polar and dispersive components as different as possible, and one of the liquids must be apolar.

The procedure was applied to determine the SFE of POM and PTFE using water and methylene iodide, which satisfied the mentioned requirements. The largest error in the total SFE calculated was always below 6%.

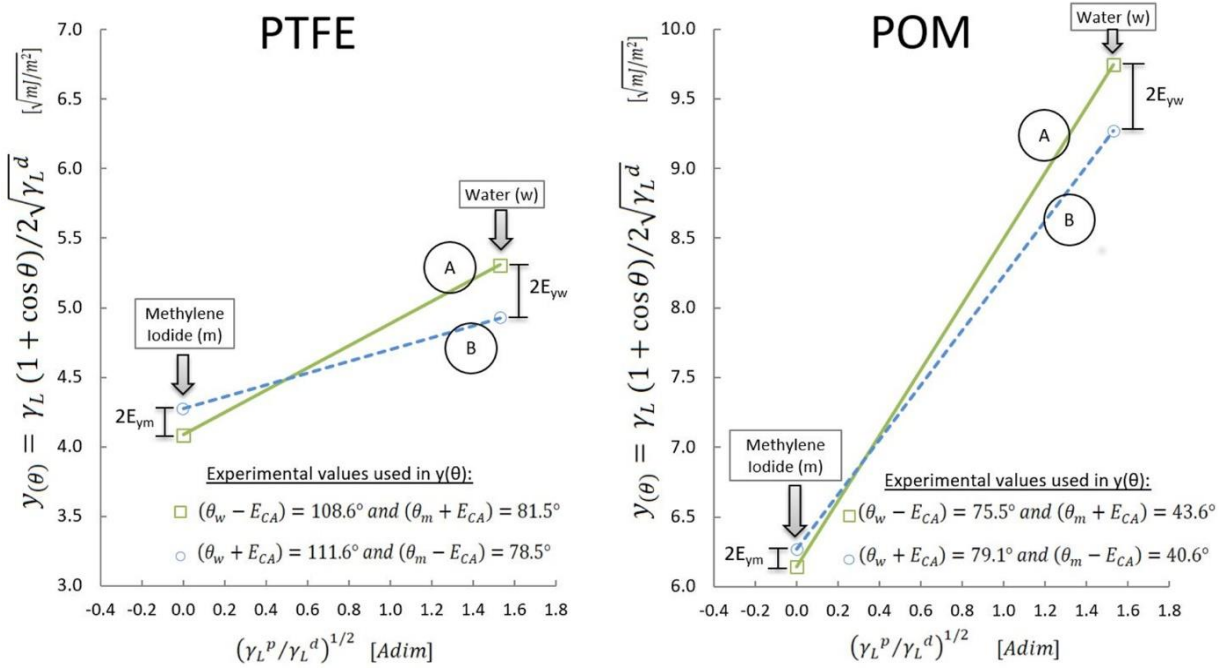

*Figure 8. Surface free energy plots for PTFE and POM. (A) Linear function (equation 9) obtained from $(\theta_w - E_{CA})$ and $(\theta_m + E_{CA})$ used to calculate $\gamma_S^p{}_{max}$ and $\gamma_S^d{}_{min}$. (B) Linear function (equation 9) obtained from $(\theta_w + E_{CA})$ and $(\theta_m - E_{CA})$ used to calculate $\gamma_S^p{}_{min}$ and $\gamma_S^d{}_{max}$. Reference: $E_{yj}$ absolute error of $y_{(\theta j)}$ (equation A.1), where $\theta_j$ is the contact angle of liquid j.*

## Appendix A. Uncertainty analysis for $y(\theta)$

In this section, we analyze the effect of the error in the determination of the contact angle on the calculation of SFE. To do this, function $y(\theta)$ is defined as the left-hand side of equation 9, i.e.:

$$y(\theta) = \gamma_L (1 + \cos\theta)/2\sqrt{\gamma_L^d} \qquad (A.1)$$

Function $y(\theta)$ is a function of the surface tension of the liquids used to measure the respective contact angles. The error in $y(\theta)$ is due to the errors committed in the measurement of the contact angles, since the values of surface tension of the liquids coming from tables are assumed to be exact. Therefore, the absolute error Ey in $y(\theta)$ can be defined as half of the interval between the values of $y(\theta)$ at $(\theta + E_{CA})$ and $(\theta - E_{CA})$ as:

$$E_y = (y(\theta + E_{CA}) - y(\theta - E_{CA}))/2 \qquad (A.2)$$

$E_y$ determines the interval of uncertainty in the calculation of $y(\theta)$ due to the absolute error in the measurement of $\theta$; these are represented in Figure 8 as $E_{yw}$ and $E_{ym}$ for each of the two solids, PTFE and POM, for water (w) and methylene iodide (m).

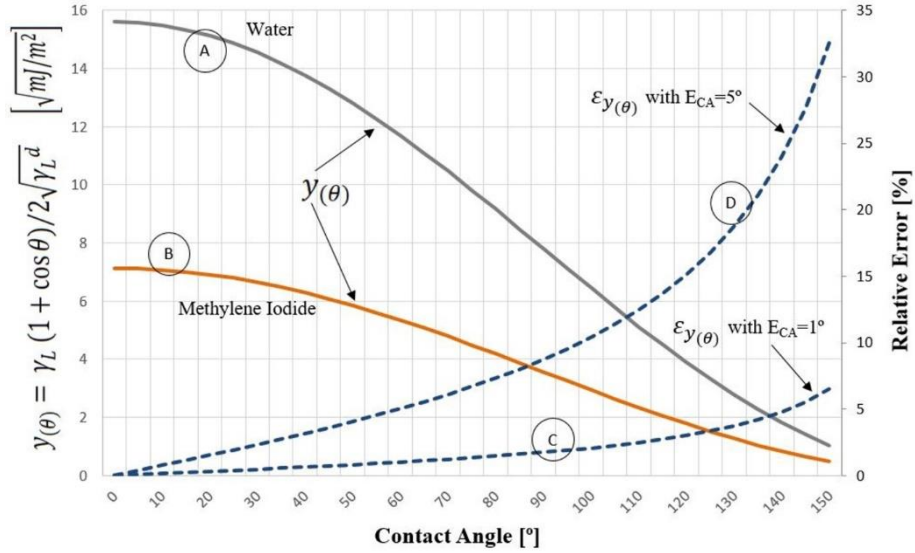

*Figure A.1. Behavior of the function $y_{(\theta)}$, between 0° and 150°, for (A) water and (B) methylene iodide. Relative error $\varepsilon_{y_{(\theta)}}$, between 0° and 150°, calculated using two absolute errors in contact angle $(E_{CA})$ of (C) 1° and (D) 5°. Observe that, according to equation A.3, the behavior of $\varepsilon_{y_{(\theta)}}$ is independent of the liquid used, $\varepsilon_{y_{(\theta)}} = \varepsilon_{CA}\, C_{(y)}$.*

Reducing the uncertainty $E_{yi}^{j}$ (i=liquid, j=solid) requires the analysis of the effect of the contact angle in $y(\theta)$ to obtain a criterion for the optimal selection of liquids for a given solid whose SFE will be determined. Figure A.1 shows the behavior of $y(\theta)$ for contact angles between 0° and 150° for water and methylene iodide where it is clearly seen that as $\theta$ increases, $y(\theta)$ decreases. Also, for any value of $\theta$, the value is larger for water than for methylene iodide, i.e. $y(\theta w) > y(\theta m)$, with a difference that decreases as $\theta$ increases. In addition, in Figure A.1, the relative errors as a function of contact angle are represented considering two scenarios of error in the determination of $\theta$ of $E_{CA} = 1$ and $E_{CA} = 5$, a range that largely exceeds the largest absolute error in this report of $E_{CA} = 1.8$, which is exclusively due to the procedure of contact angle measurement and not to the specific liquid used.

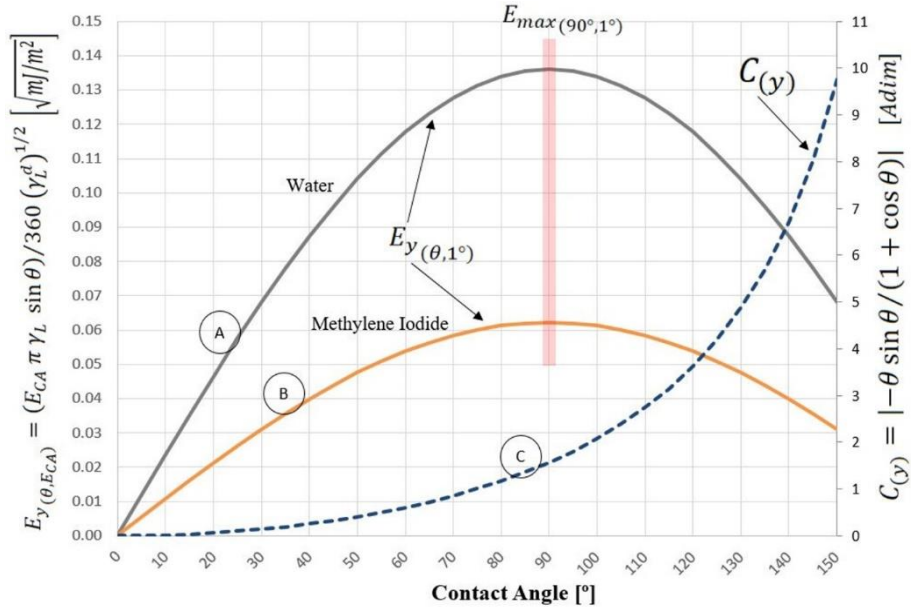

*Figure A.2. Behavior of the function $E_{y_{(\theta, E_{CA})}}$, between 0° and 150°, for (A) water and (B) methylene iodide calculated using one absolute error in contact angle ($E_{CA}$) of 1°. Plot of the (C) condition number function $C_{(y)}$.*

On the other hand, the condition number $C(y)$ of $y(\theta)$, which is the value of the relative error in the output function, i.e.: $\Delta y(\theta)/y(\theta) = \varepsilon_y$, divided by the value of the relative error in the input variable, i.e.: $\varepsilon_{CA}$, is given in equation A.3:

$$C(y) = \varepsilon_y/\varepsilon_{CA} \cong \left|\theta \left[\tfrac{dy}{d\theta}/y(\theta)\right]\right| \cong |-\theta \sin\theta/(1+\cos\theta)| \tag{A.3}$$

$C(y)$ continuously increases with $\theta$, as shown in Figure A.2. This Figure shows that $C(y) < 1$ for $\theta < 75°$, which is desirable because, in this range, the error in $y(\theta)$ is not magnified by the error in $\theta$. For $\theta > 75°$, $C(y)$ increases up to a value of 10 at $\theta = 150°$, which indicates that the error in $y(\theta)$ is magnified up to ten times due to $\varepsilon_{CA}$.

It is possible to derive a relation between the absolute error $E_{CA}$ and the absolute error of function $y_{(\theta)}$; $E_y$ as:

$$E_y(\theta, E_{CA}) = (E_{CA} \pi \gamma_L \sin\theta)/360 \left(\gamma_L^d\right)^{1/2} \tag{A.4}$$

which is plotted in Figure A.2 as a function of $\theta$ for water and methylene iodide.

### Appendix B. Uncertainty analysis for the components: $\gamma_s^p$ and $\gamma_s^d$

Recalling equation 9, the function $y(\theta)$ vs $\left(\gamma_L^p/\gamma_L^d\right)^{1/2}$ is linear, with a slope ($m_y$) which is $\sqrt{\gamma_L^p}$, that is, the square root of the polar component of the surface tension of the solid.

Therefore,

$$\gamma_s^p = ((y(\theta i) - y(\theta k))/z)^2 \tag{B.1}$$

where $\theta i$ and $\theta k$ are the values of the contact angles for liquids i and k, respectively, and z is the difference in the value of the abscissa;

$$z = \sqrt{\gamma_i^p/\gamma_i^d} - \sqrt{\gamma_k^p/\gamma_k^d} \tag{B.2}$$

The absolute error of $\gamma_s^p$ from Equation B.1 is:

$$E_{\gamma_s^p} \cong |\partial \gamma_s^p/\partial y(\theta i)\, E_{yi}| + |\partial \gamma_s^p/\partial y(\theta k)\, E_{yk}| \cong 2m_y(E_{yi} + E_{yk})/z \tag{B.3}$$

The resulting relative error is then:

$$\varepsilon_{\gamma_s^p} = E_{\gamma_s^p}/\gamma_s^p = E_{\gamma_s^p}/m_y^2 \cong 2(E_{yi} + E_{yk})/(m_y\, z) \tag{B.4}$$

As can be observed, the relative error in the determination of the polar component $\gamma_s^p$ is a function of three terms, inversely proportional to $m_y$ and z and directly proportional to $(E_{yi} + E_{yk})$. $m_y$ is $\sqrt{\gamma_s^p}$, which, as expected, for a given absolute error, the relative error decreases with the value of the measured magnitude.

In the case of z, given by equation B.2, it decreases as the ratio $\sqrt{\gamma_L^p/\gamma_L^d}$ for both liquids i and k become similar, therefore the relative error increases rapidly. As examples, let's consider two cases: one in which the liquids selected are water and another non-polar liquid such as methylene iodide, alpha-bromonaphthalene, benzene, etc., in which z will have a value of 1.53, and another in which the liquids selected are water and a less polar liquid like glycerol, ethylene glycol, etc., in which the value of z will be ~0.7 and will give an error ~120% higher than in the first case (see Table B.1 and Figure B.1).

The term $(E_{yi} + E_{yk})$ is the sum of the errors in the calculation of $y(\theta_{i,k})$ due to the error in the contact angle measurement. According to equation A.4, these errors have maximum values for a contact angle of $\theta = 90°$ for any absolute error $E_{CA}$ in the measurement of the contact angle. This value is intrinsically associated with the precision of the measurement procedure and should be kept to a minimum value. It must be noted that the value of the polar component of the surface free energy, and therefore of the total surface free energy, is more sensitive to the errors produced in the measurement of the contact angel of water than the errors produced in the contact angle of methylene iodide.

*Table B.1. Values of the ratio between polar and dispersive components for different liquids* [2], [6].

| | Water (w) | Glycerol (g) | Ethylene glycol (e) | Formamide (f) | Dimethyl sulfoxide (d) | Methylene iodide (m) |
|---|---|---|---|---|---|---|
| $\gamma_L^p$ [mJ/m²] | 51.00 | 30.00 | 19.00 | 19.00 | 8.00 | 0.00 |
| $\gamma_L^d$ [mJ/m²] | 21.80 | 34.00 | 29.00 | 39.00 | 36.00 | 50.80 |
| $\sqrt{\gamma_L^p/\gamma_L^d}$ | 1.53 | 0.94 | 0.81 | 0.70 | 0.47 | 0.00 |

In the case of the dispersive component of the SFE, its square root represents the ordinate to the origin of the linear function given by equation 9, of $y(\theta)$ vs $\sqrt{\gamma_L^p/\gamma_L^d}$, that is:

$$\sqrt{\gamma_s^d} = b = -\sqrt{\gamma_k^p/\gamma_k^d} \cdot m + y(\theta k) \tag{B.5}$$

For a given liquid k, which can be rewritten as:

$$\gamma_s^d = b^2 = \left(t(y(\theta i) - y(\theta k)) + y_{(\theta k)}\right)^2 \tag{B.6}$$

where *t* is:

$$t = -\sqrt{\gamma_k^p/\gamma_k^d}/z = -\sqrt{\gamma_k^p/\gamma_k^d}/\left(\sqrt{\gamma_i^p/\gamma_i^d} - \sqrt{\gamma_k^p/\gamma_k^d}\right) \tag{B.7}$$

The absolute error in $\gamma_s^d$ by definition is:

$$E_{\gamma_s^d} \cong \left|\partial\gamma_s^d/\partial y(\theta i)\ E_{yi}\right| + \left|\partial\gamma_s^d/\partial y(\theta k)\ E_{yk}\right| \cong 2b\left(|t|E_{yi} + |1-t|E_{yk}\right) \tag{B.8}$$

and the relative error in $\gamma_s^d$ is therefore:

$$\varepsilon_{\gamma_s^d} = E_{\gamma_s^d}/\gamma_s^d = E_{\gamma_s^d}/b^2 \cong 2\left(|t|E_{yi} + |1-t|E_{yk}\right)/b \tag{B.9}$$

First, it is observed that, if the liquid is non-polar, i.e. $\gamma_i^p \cong 0$, and then $\sqrt{\gamma_i^p/\gamma_i^d} \cong 0$, the value of t=1, and the value of $\gamma_s^d$ given by equation B.6 is reduced to:

$$\gamma_s^d = b^2 = y(\theta i)^2 \tag{B.10}$$

with absolute and relative errors given by equations B.11 and B.12 respectively:

$$E_{\gamma_s^d} \cong \left|\partial\gamma_s^d/\partial y(\theta i)\ E_{yi}\right| \cong 2\ y(\theta i)\ E_{yi} \tag{B.11}$$

$$\varepsilon_{\gamma_s^d} = E_{\gamma_s^d}/\gamma_s^d = E_{\gamma_s^d}/y(\theta i)^2 \cong 2\ E_{yi}/y(\theta i) \tag{B.12}$$

In the case of a polar liquid, the relative error in the determination of the dispersive component, given by equation B.9, is inversely proportional to b, which is the error of the ordinate to the origin calculated with equation B.6 and directly and directly proportional to $\left(|t|E_{yi} + |1-t|E_{yk}\right)$, where t is given by equation B.8. In such case, the value of t increases as the value of $\sqrt{\gamma_i^p/\gamma_i^d}$ gets closer to the value of $\sqrt{\gamma_k^p/\gamma_k^d}$, that is, the liquids used to determine the contact angles have similar ratios between the polar and dispersive components of SFE. Therefore, in such cases, the relative error in the determination of $\gamma_s^d$ will be either very large, as in the case of glycerol and ethylene glycol or moderately large, as in the case of water and formamide (Figure B.1). As conclusion of this analysis, the relative error in each component $\gamma_s^p$ and $\gamma_s^d$ of SFE shows that the relative errors are reduced if the selection of liquids satisfies the following;

-the liquids used must have different values of the ratio between polar and dispersive components.

-one of the liquids should be apolar.

-the precision for the procedure for contact angle measurement must be as high as possible.

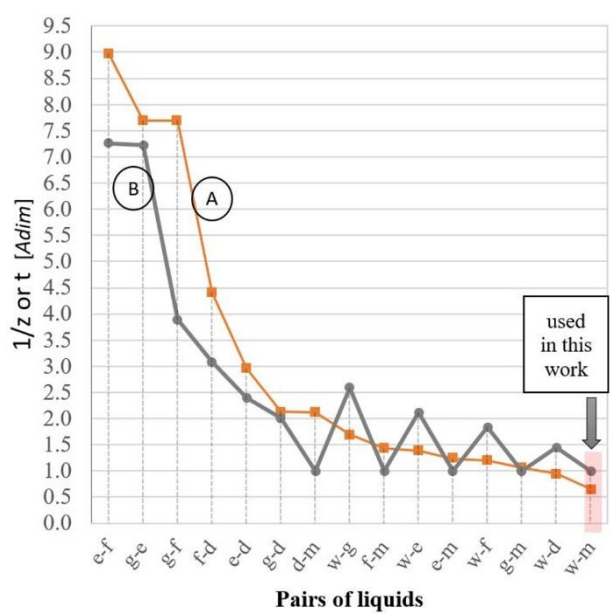

*Figure B.1. Values of 1/z (A) and t (B) for different pairs of liquids. The relationship between these values, $\varepsilon_{\gamma_s^p}$ and $\varepsilon_{\gamma_s^d}$ is described in equations B.4 and B.9. References: see Table B.1.*

**Acknowledgements**

We thank Cristian Cegelski for help in the construction of the goniometer. Funding: This work was supported by Universidad Nacional de Misiones [Proyectos de Investigación con Impacto Tecnológico y Social 2015/2016, A-15, 16/Q589], Argentina.